\documentclass[letter]{mn2e}
\usepackage{psfig}
\newcommand{\ltsima} {$\; \buildrel < \over \sim \;$}
\newcommand{\gtsima} {$\; \buildrel > \over \sim \;$}
\newcommand{\lta} {\lower.5ex\hbox{\ltsima}}
\newcommand{\gta} {\lower.5ex\hbox{\gtsima}}

\newcommand{\Ha}{H$\alpha$}

\newcommand{\Lya}{Ly$\alpha$}
\title [The emission halo of LAB1]
{Dissecting the Lyman $\alpha$ emission halo of LAB1}

\author[A. Weijmans et al.]
  {Anne-Marie Weijmans$^{1,2}$\thanks{E-mail: weijmans@di.utoronto.ca}, Richard G. Bower$^{3}$, James E. Geach$^{3}$, \and  A. Mark Swinbank${^3}$, R.~J. Wilman$^{4}$, P.~T. de Zeeuw$^{5,1}$ and Simon L. Morris$^{3}$ \\
  $^1$ Sterrewacht Leiden, Leiden University,  Postbus 9513, 2300 RA
  Leiden, The Netherlands  \\
  $^2$ Dunlap Institute for Astronomy \& Astrophysics, University of
  Toronto, 50 St. George Street, Toronto, ON M5S 3H4, Canada \\
  $^3$ Institute for Computational Cosmology, Department of Physics, Durham University, South Road, Durham, DH1 3LE, UK \\
  $^4$ Centre for Astrophysics \& Supercomputing, Swinburne University of Technology, Hawthorn, Victoria 3122, Australia \\
  $^5$ European Southern Observatory, Karl-Schwarzschild-Str 2, 85748 Garching, Germany \\ }

\pagerange{\pageref{firstpage}--\pageref{lastpage}} \pubyear{2009}

\begin{document}

\maketitle

\label{firstpage}

\begin{abstract} 
  We report observations of Lyman Alpha Blob 1 (LAB1) in the SSA 22
  protocluster region ($z=3.09$) with the integral-field spectrograph
  SAURON. We increased the signal-to-noise in the spectra by more
  than a factor three compared to our previous observations. This
  allows us to probe the structure of the LAB system in detail,
  examining its structure in the spatial and wavelength dimensions. We
  find that the emission from the system comes largely from five
  distinct blobs. Two of the emission regions are associated with
  Lyman Break Galaxies, while a third appears to be associated with a
  heavily obscured submillimeter galaxy. The fourth and fifth
  components do not appear to be associated with any galaxy despite
  the deep imaging that is available in this field. If we interpret
  wavelength shifts in the line centroid as velocity structure in the
  underlying gas, many of these emission systems show evidence of
  velocity shear. It remains difficult to distinguish between an
  underlying rotation of the gas and an outflow driven by the central
  object.  We have examined all of the line profiles for evidence of
  strong absorption features. While several systems are better fitted
  by the inclusion of a weak absorption component, we do not see
  evidence for a large-scale coherent absorption feature such as that
  seen in LAB2.

    \date{}
\end{abstract}



\section{Introduction}
\label{sec:introduction}

By allowing us to probe the gaseous haloes around $z \sim 3$ galaxies,
large scale Lyman $\alpha$ nebulae provide a fascinating insight into
the formation of high-redshift galaxies. The first and brightest of
these haloes were discovered by Steidel et
al. (2000)\nocite{2000ApJ...532..170S} in the SSA 22 protocluster
region at $z=3.09$. Subsequently a population of fainter Lyman Alpha
Blobs (LABs) was detected in deep narrow-band imaging surveys
(e.g. Matsuda et al. 2004\nocite{2004AJ....128..569M}; Nilsson et al.
2006\nocite{2006A&A...452L..23N}, Smith \& Jarvis
2007\nocite{2007MNRAS.378L..49S}), revealing that LABs have a large
spread in properties such as surface brightness and morphology, and it
has been suggested that their presence is linked to dense environments
(Matsuda et al.  2004\nocite{2004AJ....128..569M}). Follow-up
observations in the optical, near-infrared and particularly the
far-infrared suggest that LABs are sites of massive galaxy formation,
enhanced by the cluster environment (e.g. Chapman et
al. 2004\nocite{2004ApJ...606...85C}).  This view is supported by the
discovery of luminous submillimeter sources in several LABs
(e.g. Geach et al. 2005\nocite{2005MNRAS.363.1398G}). LAB systems like
those discovered by Steidel et al. (2000)\nocite{2000ApJ...532..170S}
are also found around high-redshift radio galaxies
(e.g. Chambers, Miley \& van Breugel 1990\nocite{1990ApJ...363...21C};
Villar-Mart\'in et al. 2002\nocite{2002MNRAS.336..436V}). However, the
study of such radio loud systems is complicated by the presence of
radio jets and lobes.  It is unclear whether these systems are
directly comparable to the radio quiet LABs that we discuss in this
paper, or that they have a different power source, such as the
injection of cosmic rays by the radio source (e.g. Ferland et
al. 2009\nocite{2009MNRAS.392.1475F}).

The origin of radio quiet LABs is still unclear, and three different
scenarios have been proposed to explain their existence. One is that
the gas in LABs is heated by photo-ionisation, caused by massive stars
and/or active galactic nuclei (AGN) (Geach et
al. 2009\nocite{2009ApJ...700....1G}). However, for one third of the
LABs in the sample of Matsuda et
al. (2004)\nocite{2004AJ....128..569M}, the observed UV luminosities
are too low to produce the observed \Lya\ radiation, although putative
ionising sources could be obscured along our line of sight. For
example, Geach et al. (2009)\nocite{2009ApJ...700....1G} argue that in
a large fraction of LABs there is sufficient UV flux from an obscured
AGN component to power the extended line emission, despite large dust
covering fractions. While photo-ionisation is likely to play a role in
powering LABs, it is nonetheless instructive to consider alternative
or additional power sources for the \Lya\ emission in LABs, since it
is not clear that a single mechanism (such as photo-ionisation) is
responsible for all of the observed properties of these objects.

An alternative scenario is that the gas in LABs is excited by cooling
flows (e.g. Haiman, Spaans \& Quataert
2000\nocite{2000ApJ...537L...5H}; Fardal et al.
2001\nocite{2001ApJ...562..605F}; Dijkstra \& Loeb
2009\nocite{2009arXiv0902.2999D}). Nilsson et
al. (2006)\nocite{2006A&A...452L..23N} argue that the emission in the
LAB they detected in the GOODS South field originates from cold
accretion onto a dark matter halo. This view is supported by the lack
of continuum counterparts, which could photo-ionise the gas, and the
absence of a massive starburst in the infrared, that could point to a
superwind outflow (see below). Instead, they find a good match between
their observed surface brightness profile and the profiles derived
from theoretical models for collapsing clouds from Dijkstra, Haiman \&
Spaans (2006)\nocite{2006ApJ...649...14D}. A similar analysis is given
by Smith et al. (2008)\nocite{2008MNRAS.389..799S} for the LAB
presented in Smith \& Jarvis (2007)\nocite{2007MNRAS.378L..49S}.

The third proposed origin for extended \Lya\ emission haloes is
provided by the so-called superwind model (e.g. Taniguchi \& Shioya
2000\nocite{2000ApJ...532L..13T}; Ohyama et al.
2003\nocite{2003ApJ...591L...9O}). After an initial starburst, massive
stars die in supernovae. If the resulting supernova remnants overlap,
they could form a superbubble (e.g. Heckman, Armus \& Miley
1990\nocite{ 1990ApJS...74..833H}), from which a superwind can blow gas
into the intergalactic medium if the kinetic energy in the gas is
large enough to overcome the gravitational potential. Taniguchi \&
Ohyama (2000)\nocite{2000ApJ...532L..13T} suggest an evolutionary
sequence for elliptical galaxies that includes LABs. During the initial
starburst, a galaxy can be enshrouded by gas and dust grains, and is
therefore observable as a (dusty) submillimeter source.  The LABs
represent the subsequent superwind phase, expelling the gas and dust,
and therefore rendering the galaxy fainter in the submillimeter
regime. After this phase, the galaxy continues to develop into a
normal elliptical galaxy.

The brightest and most extended LAB observed to date is LAB1
(SSA22a-C11), one of the two LABs described by Steidel et al.
(2000)\nocite{2000ApJ...532..170S}. This LAB has a \Lya\ luminosity of
$1.1 \times 10^{44}$ ergs s$^{-1}$ at $z=3.1$ (Matsuda et al.
2004\nocite{2004AJ....128..569M}), and a spatial extent of $\sim$ 100
kpc. Matsuda et al. (2004)\nocite{2000ApJ...532..170S} find
bubble-like structures in narrowband images of LAB1, in support of the
superwind model for this LAB. Bower et
al. (2004)\nocite{2004MNRAS.351...63B} observed LAB1 with the
integral-field spectrograph SAURON (Bacon et
al. 2001\nocite{2001MNRAS.326...23B}). They found extensive emission, and
a large velocity dispersion for the \Lya\ emission line ($\sim$ 500 km
s$^{-1}$). Although interpretation of the spectra of LAB1 is not
straightforward, since \Lya\ is a resonant line and therefore emission
can easily get scattered, they find similarities between LAB1 and the
local emission-line halo of NGC 1275 in the Perseus cluster. Longer
observations of LAB2 with the same spectrograph suggested the
existence of a dense outflowing shell of material around this system
(Wilman et al. 2005\nocite{2005Natur.436..227W}). However, spectra of
higher signal-to-noise are needed to find evidence for outflows in
LAB1.

We therefore reobserved LAB1 with SAURON, adding signal to the data
already published by Bower et al. (2004)\nocite{2004MNRAS.351...63B}.
By increasing the observing time from 9 hours in the original dataset
to 23.5 hours in our new datacube, we increased the signal-to-noise in
the spectra, and therefore were able to obtain line profiles and
kinematic maps of the \Lya\ emission in this region. We describe the
new observations and data reduction of LAB1 in
Section~\ref{sec:observations} and analyse the spectra in
Section~3. In Section 4 we discuss our results and speculate on the
structure and origin of LAB1. Throughout this paper, we assume a flat
cosmology with $H_0 = 70$ km s$^{-1}$ Mpc$^{-1}$, $\Omega$=0.3 and
$\Lambda$ = 0.7. In this scenario, 1 arcsec at $z=3.1$ corresponds to
7.5 kpc.


\begin{figure*}
\begin{center}
\psfig{figure=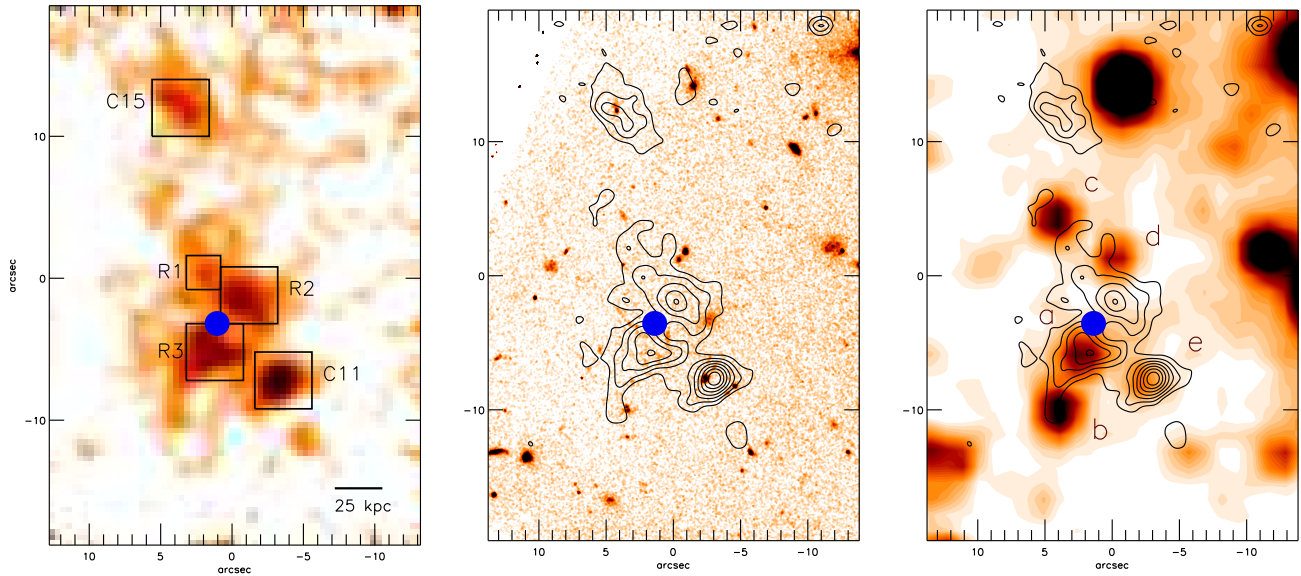,angle=90,width=18cm}
\caption{\Lya\ emission in the LAB1 region. Left panel: continuum
subtracted \Lya\ emission, obtained from collapsing the SAURON spectra
over a narrow wavelength range centred on the emission
line. Interesting regions are indicated by boxes (see text). Middle
panel: {\it HST}/STIS optical image overlaid with \Lya\ contours from
the left panel. The faintest contour has a surface brightness of
5.6 $\times$ 10$^{-18}$ erg s$^{-1}$ cm$^{-2}$ arcsec$^{-2}$ and
contours increase in steps of 3.7$\times$ 10$^{-18}$ erg s$^{-1}$
cm$^{-2}$ arcsec$^{-2}$. Note however the 50 per cent uncertainty in
absolute flux calibration (see \S2). Right panel: same as
middle panel, but now a {\it Spitzer}/IRAC 3.6 $\mu$m image is
displayed. Sources identified in Geach et al. (2007) are indicated
with identical nomenclature ($a$-$e$). In all plots, the blue dot
denotes the position of the radio source (Chapman et al. 2004). All
images are plotted on the same scale and are orientated such that
North is up and East to the left.}
\label{fig:lya}
\end{center}
\end{figure*}

\section{Observations and data reduction}
\label{sec:observations}

LAB1 was previously observed for 9 hours with SAURON at the William
Herschel Telescope at La Palma, Spain, in July 2002. The relatively
large field-of-view (41 $\times$ 33 arcsec$^2$) and high throughput
(20 per cent) make this spectrograph a very suitable instrument for
deep \Lya\ observations, even though it was originally built to study
the dynamics and stellar populations of nearby early-type galaxies (de
Zeeuw et al. 2002\nocite{2002MNRAS.329..513D}). The relatively high
spectral resolution of 4.2 \AA\ (FWHM) of the instrument, in
combination with the wide field, was obtained by compromising on the
total wavelength coverage (4810 to 5350 \AA). As a result, SAURON can
observe \Lya\ systems at redshifts $2.96 < z < 3.40$, and fortunately,
LAB1 resides within this redshift range ($z=3.1$). A description of
the earlier observations of LAB1 and their reduction can be found in
Bower et al. (2004)\nocite{2004MNRAS.351...63B}.

In these previous observations, we lacked the signal-to-noise ($S/N$)
to search for asymmetries in the line profiles and signs for possible
neutral absorption, as found in SAURON spectra of LAB2 by Wilman et
al. (2005)\nocite{2005Natur.436..227W}. We therefore reobserved LAB1
for an additional 15 hours with SAURON, between 15 and 21 September
2006. The observations were split into individual exposures of 1800
seconds and dithered by a few arcseconds. The data were reduced using
the dedicated \textsc{xsauron} software (Bacon et
al. 2001\nocite{2001MNRAS.326...23B}). We replaced our calibration
frames, to correct for a malfunctioning shutter (see Weijmans et
al. 2009\nocite{2009MNRAS.398..561W} for more details). We had also
observed six blank sky fields, and created a so-called superflat by
taking the smoothed median of these blank frames and the dithered
object frames in the spectral direction. Dividing our already
flat-fielded data by this superflat removed most of the remaining
flat-field residuals. The spectra were degraded to a resolution of 4.9
\AA\ (FWHM) to be consistent with the LAB1 observations of Bower et
al. (2004)\nocite{2004MNRAS.351...63B}, and sky subtracted with the
signal from the SAURON skylenslets, that obtain simultaneous
sky spectra pointing 2 arcminutes away from the main field-of-view.

Before merging this new dataset with the cubes of Bower et
al. (2004)\nocite{2004MNRAS.351...63B}, we first re-reduced these
spectra in the same way as described above to make sure that both
datasets were treated identically. One of the frames had to be
discarded because of bad sky subtraction. The remaining 17 frames of
the old dataset were then merged with the 30 frames of the new
dataset, using a faint star in the south-east corner of the
field-of-view to align the cubes. In the merged cube, this star has a
FWHM of 1.3 arcsec. We set the spatial resolution of the cube to 0.4
arcsec per pixel, while the spectral resolution is 1.15 \AA\ per
pixel. This final cube represents 23.5 hours of observing time, and is
the deepest SAURON observation to date. The increase in $S/N$ is a
factor of 3.7 compared to the dataset of Bower et
al. (2004)\nocite{2004MNRAS.351...63B}, which is more than would be
expected based on the factor 2.6 increase in exposure time. The higher
than expected $S/N$ results from improved observing and data reduction
techniques, developed over the past few years within the SAURON team.

Unfortunately, due to the malfunctioning shutter of the spectrograph we
could not observe flux stars during the 2006 observing run. 
However, since relative run-to-run variations in spectral response of
SAURON have been shown to be smaller than 1 per cent (Kuntschner et
al. 2006\nocite{2006MNRAS.369..497K}), we could still performe a
relative flux calibration using flux star observations from a previous
run. An absolute flux calibration of our data was done using flux
values for C15 from the literature. We added the flux in a
wavelength interval centred around the \Lya\ line (4960 \AA\ $<$
$\lambda$ $<$ 5000 \AA) in a 4 $\times$ 4 arcsec$^2$ box around C15,
and scaled the number of counts to the flux given by Matsuda et
al. (2004)\nocite{2004AJ....128..569M}. We checked our conversion
factor with Bower et al. (2004)\nocite{2004MNRAS.351...63B}, and found
a deviation of 50 per cent. As the flux calibration is the main source
of uncertainty, we adopt this deviation to estimate the errors on our
obtained fluxes (see Table~\ref{tab:blobs}).


\section{Analysis}

\subsection{The halo structure of LAB1}
\label{sec:sub}

In Figure~\ref{fig:lya} we show a continuum subtracted \Lya\ image of
LAB1, obtained by integrating the spectra in our SAURON datacube over
a small wavelength range (4960 - 5040 \AA) containing the redshifted
emission line.  The most striking result is that LAB1 is not one
coherent structure, but that the emission is concentrated in five
distinct emission regions (labeled R1-R3, C11, C15), embedded in a
\Lya\ emission halo (see also Matsuda et
al. 2004\nocite{2004AJ....128..569M}). These five regions were
selected primarily by eye, exceeding the third contour level in
Figure~\ref{fig:lya}, corresponding to 1.3 $\times$ 10$^{-17}$ erg
s$^{-1}$ cm$^{-2}$ arcsec$^{-2}$. We estimate that about 55 per cent
of the total \Lya\ emission can be associated with one of these
regions. Thus it appears that the giant emission halo results from a
combination of smaller emission blobs, comparable to the more
typical LABs identified by Matsuda et
al. (2004)\nocite{2004AJ....128..569M}.

In order to attempt to identify each of the \Lya\ blobs (R1-R3, C11
and C15) with underlying galaxies, we overlay the \Lya\ emission line
contours of our cube with an {\it HST}/STIS image and a {\it
Spitzer}/IRAC 3.6 $\mu$m image (see Figure~\ref{fig:lya}). Geach et
al. (2007)\nocite{2007ApJ...655L...9G} used the same images as well as
MIPS 24 $\mu$m imaging to identify IRAC counterparts in the LAB1
region. They found five sources, labeled $a$-$e$ in
Figure~\ref{fig:lya}, of which two ($c$ and $d$) have mid-infrared
colours inconsistent with galaxies at $z = 3.1$, and therefore are
most likely not part of the proto-cluster. The IRAC source $b$ is
located at the southern extreme of LAB1, and the new SAURON
observations reveal that it is not coincident with a peak in the
Ly$\alpha$ emission (although there is some low-surface brightness
emission extending from the north). The two remaining IRAC sources,
$a$ and $e$, seem to be related to the \Lya\ emission. We discuss the
\Lya\ emission blobs in more detail below.

Two of the brightest \Lya\ emitting regions (C11 and C15) were
identified with Lyman Break Galaxies (LBGs) by Steidel et
al. (2000)\nocite{2000ApJ...532..170S}. These can be clearly seen in
the STIS continuum image. The positional uncertainty between the \Lya\
emission and continuum detection in the STIS image is less than 1.5
arcsec.  C11 is also weakly detected in the IRAC bands (source $e$),
with a 3.6 $\mu$m flux of 1.5 $\mu$Jy. C15 was labeled by Matsuda et
al. (2004)\nocite{2004AJ....128..569M} as LAB8, as this blob is
clearly separated from the main emission halo. Neither C15 nor C11 is
detected in X-ray ($L_X < 2.0 \times 10^{43}$ ergs s$^{-1}$, 2-43 keV
band) in the deep {\it Chandra} exposures of this region (Geach et
al. 2009\nocite{2009ApJ...700....1G}).

The bright Ly$\alpha$ blob R3 is associated with an extremely red
galaxy (source $a$ in Geach et al. 2007\nocite{2007ApJ...655L...9G})
and thought to be the counterpart to a bright submillimeter source
detected by Chapman et al. (2001\nocite{2001ApJ...548L..17C};
2004\nocite{2004ApJ...606...85C}). This source has an unresolved
submillimeter flux of S$_{850 \mu \mathrm {m}}$ = 16.8 $\pm$ 2.9 mJy
and S$_{450 \mu \mathrm{m}}$ = 45.1 $\pm$ 15.5 mJy, measured with the
Submillimeter Common-User Bolometer Array (SCUBA). The peak of the
Ly$\alpha$ emission of R3 lies within 1.3 arcsec of the submillimeter
centroid, and 0.7 arcsec of the IRAC counterpart $a$. Higher
resolution submillimeter observations with the Submillimeter Array
(SMA) (Matsuda et al. 2007)\nocite{2007ApJ...667..667M} yielded no
detection, and hinted that the submillimeter emission could originate
from an extended starburst component on scales of $>$4 arcseconds
($>$30\,kpc), and indeed this source is co-incident with several
low-surface brightness UV components in the STIS imaging.

Nearby tentative radio and CO detections (Chapman et
al. 2004\nocite{2004ApJ...606...85C}) further re-inforce the view that
the submillimeter source is associated with the IRAC source $a$, and
therefore most likely with \Lya\ blob R3. Because of the large
positional uncertainties on the radio detection, which could be
consistent with either the submillimeter source or the cavity in the
\Lya\ emission that separates R3 from R2, we prefer to associate the
radio source with the submillimeter source, given the strong
association between those identifications in other submillimeter
studies (e.g. Ivison et
al. 2007\nocite{2007MNRAS.380..199I}). Assuming a modified blackbody
spectrum with characteristic temperature $T_d$ = 35\,K (Blain, Barnard
\& Chapman 2003\nocite{2003MNRAS.338..733B}), the observed 850\,$\mu$m
flux corresponds to a bolometric luminosity of $L_{\mathrm{bol}}$ =
1.5 $\times$ 10$^{13}$ $L_\odot$. If this luminosity arises from
starformation alone with a standard IMF (Kennicutt
1998\nocite{1998ApJ...498..541K}), then the implied star formation
rate (SFR) is $\sim$ 2500\,$M_\odot$ yr$^{-1}$. We note that an AGN
contribution in LAB1 is not likely to significantly affect this
result: LAB\,1 is not detected in a 400\,ks {\it Chandra} exposure,
with a luminosity limit $L_X < 2.4 \times 10^{43}$ ergs s$^{-1}$ in
the 2--32 keV band (Geach et
al. 2009\nocite{2009ApJ...700....1G}). The most likely scenario is
that source $a$ is dominated by a dusty, potentially extended,
starburst.

In contrast to C11, C15 and R3, the \Lya\ emission regions R1 and R2
do not appear to have either an optical or a mid-infra red counterpart
(down to a 3.6 $\mu$m flux of $< 1 \mu$Jy). Given the depth of the
IRAC imaging of this field it seems unlikely that these sources are
identified with a dust obscured galaxay. Although offsets of several
arcseconds are also seen in the field \Lya\ emission survey of Matsuda
et al. (2004)\nocite{2004AJ....128..569M}, analysis of the optical to
mid-IR colours suggests that the nearest IRAC source $d$ is not part
of the proto-cluster system.  An appealling possibility is therefore
that components R1 and R2 are genuinely associated with gas
trapped, and possibly cooling, within the proto-cluster potential. R1
and R2 are at the same velocity as the other components in LAB1 and
show no clear signs of outflow (see \S\ref{sec:kin}) or association
with continuum sources. This highlights the diverse range of physics
that could be contributing to the overall \Lya\ emission structure,
which is perhaps powered by a combination of feedback from young
active galaxies {\it and} energy released from the cooling of pristine
gas.

\begin{table*}
\begin{center}
\begin{tabular}{l|c|c|c|c|c|}
\hline\hline Field & $\alpha$ & $\delta$ & L (\Lya) & 
STIS & $m_{3.6\mu m}$ \\ 
 & &  & erg s$^{-1}$ & mag & mag  \\
\hline 
C15 & 22$^{\mathrm{h}}$ 17$^{\mathrm{m}}$ 26.1$^{\mathrm{s}}$ &
00$^\circ$ 12$^{\prime}$ 54.1$^{\prime\prime}$ &  1.7 $\times$ 10$^{43}$ &  26.13 $\pm$ 0.02  & $>$23.7 \\ 

R1 & 22$^{\mathrm{h}}$ 17$^{\mathrm{m}}$ 26.0$^{\mathrm{s}}$ & 00$^\circ$
12$^{\prime}$ 42.5$^{\prime\prime}$ & 8.4
$\times$ 10$^{42}$ & $>$27.42 & $>$23.7 \\ 

R2 & 22$^{\mathrm{h}}$
17$^{\mathrm{m}}$ 25.8$^{\mathrm{s}}$ & 00$^\circ$ 12$^{\prime}$
40.9$^{\prime\prime}$ &  2.3 $\times$
10$^{43}$ &  $>$27.42 & $>$23.7 \\ 

R3 & 22$^{\mathrm{h}}$ 17$^{\mathrm{m}}$
26.0$^{\mathrm{s}}$ & 00$^\circ$ 12$^{\prime}$ 36.9$^{\prime\prime}$ &
 2.7 $\times$ 10$^{43}$ &  $>$27.42 & 22.6 $\pm$ 0.1 \\

C11 & 22$^{\mathrm{h}}$ 17$^{\mathrm{m}}$ 25.6$^{\mathrm{s}}$ &
00$^\circ$ 12$^{\prime}$ 34.9$^{\prime\prime}$ & 
 2.3 $\times$ 10$^{43}$ & 24.21 $\pm$ 0.01  & 23.6 $\pm$ 0.1 \\ \hline
\end{tabular}
\caption{\Lya\ blobs indicated in Figure~\ref{fig:lya}. Coordinates
are provided in J2000 notation. The uncertainties in the observed
\Lya\ luminosity are around 50 per cent (see \S2). The last two columns
give the optical ({\it HST}/STIS) and infrared ({\it Spitzer}/IRAC
3.6$\mu$m) fluxes in AB magnitudes. The IRAC fluxes have been measured
in circular apertures of 4 arcsec diameter, and corrected for a small
aperture loss. Lower magnitude limits for non-detections are 3$\sigma$ limits.}
\label{tab:blobs}
\end{center}
\end{table*}

\subsection{Emission line profiles}
\label{sec:profile}

\begin{figure}\centerline{\psfig{figure=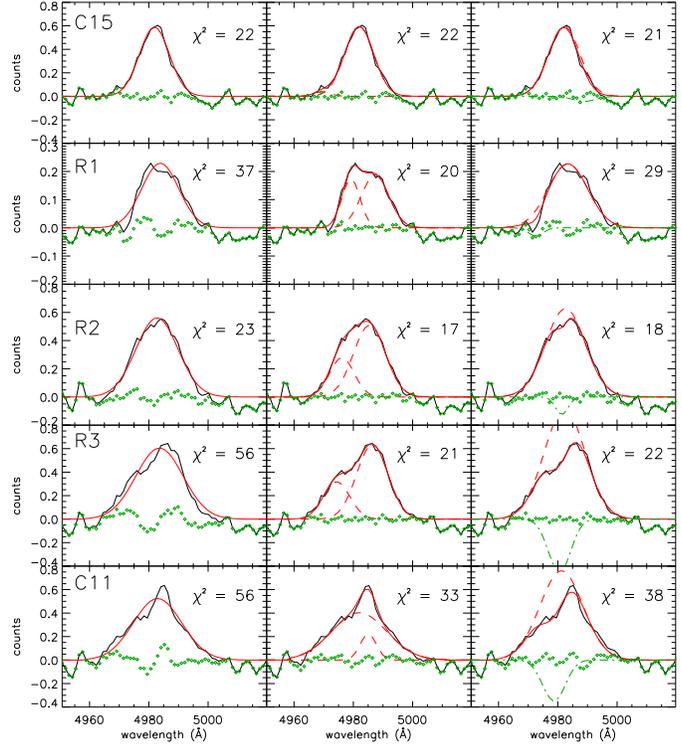,angle=0,width=9cm}}
\caption{Single Gaussian (left), double Gaussian (middle) and combined
Gaussian emission and Voigt absorption profiles (right) fitted to the
spectra in indicated regions of Figure~\ref{fig:lya}. Rows show the
spectra and fits for different blobs, as indicated in the upper left
corner of the left panel. The observed spectra are shown in black,
while the fit is overplotted in red. Dashed red lines show Gaussian
profiles, and in the right hand panels the green dotted-dashed line indicates
the absorber. The green dots show the residuals of the fit. All fits
have been convolved with the instrumental dispersion profile of 108 km
s$^{-1}$. See Table~\ref{tab:gausses} for the fitted parameters.}
\label{fig:gausses}
\end{figure}

One of the principle goals of our deeper observations was to analyse
the structure of LAB1 as a function of wavelength in search of
outflows and coherent absorption line systems, such as seen in LAB2
(Wilman et al. 2005\nocite{2005Natur.436..227W}). To analyse the line
profiles, we bin the spectra of each blob in a square region centred
on the emission peak. Each square has sides of 4 arcsec to
enclose the separate blobs, except for the smaller blob R1, where we
used a square with sides of 2.4 arcsec (Figure~\ref{fig:lya}). We
initially fit the resulting spectra with a single Gaussian emission
line, but with the exception of C15 and R2 this provides a poor fit to
the data. Better fits are obtained by instead modeling the line shape
with a double Gaussian or a Gaussian emission profile combined with a
Voigt absorption profile (for details of the fitting procedure see
Wilman et al. 2005\nocite{2005Natur.436..227W}).  In the case of
absorption, the wavelength of the underlying Gaussian profile and
Voigt profile are allowed to vary freely, but we found that if we
allowed both the column density and equivalent width of the absorption
to vary, the solutions were too degenerate because of their position
on the curve of growth. We therefore fixed the column density at
3$\times$10$^{14} \,\hbox{cm}^{-2}$, typical of the range of the fits
when this parameter was left free.

As Figure~\ref{fig:gausses} shows, a clear improvement (a change in
$\chi^2$ larger than $3\sigma$) for R1, R3 and C11 is obtained by
including the absorber, or by allowing multiple Gaussian
components. For R2, the fit improves, but the improvement in $\chi^2$
is not significant. These fits cannot, however, discern between a
situation with multiple emission sources or an absorbing medium.
Moreover, the double Gaussian fits can be a mere reflection of the
underlying velocity field of each source, as we sum spectra over
relatively large apertures. We will explore this in more detail in
\S\ref{sec:kin}.

In Table~\ref{tab:gausses} we show the parameters of the best fits. It
is remarkable that the redshift of the underlying Gaussian emission
lines between the various components is $z$ = 3.099 $\pm$ 0.001 and
varies in the restframe by less than 150 km s$^{-1}$ over the whole
region. This supports the interpretation of the LAB1 system as a
high-redshift virialised group.  The Gaussian line-widths are $\sigma
\sim$400 km s$^{-1}$, measured in the restframe of the cluster, which
is a typical value for LABs. In contrast to the case in LAB2 (Wilman
et al. 2005\nocite{2005Natur.436..227W}) we do not find that the
absorption is particularly strong, nor is the redshift of the absorber
constant across the system. In LAB2 we used this to argue for the
existence of a large-scale absorbing shell of outflowing material. In
LAB1 we find no evidence for such a feature: indeed the strength of
these putative absorption systems is such that they might well arise
in the Lyman-alpha forest surrounding the LAB1 system (Wilman et
al. 2004\nocite{2004MNRAS.351.1109W}). Rest-frame UV spectroscopy
of C11 presented by Shapley et al.  (2006)\nocite{2006ApJ...651..688S}
suggests velocity offsets  of 380 km s$^{-1}$ between the \Lya\
emission and the ISM. This is larger, but quantitatively similar to the
offset of 220 km s$^{-1}$ that we derive between the peak of the \Lya\
emitter and the blue-shifted absorber, thus suggesting that the absorber
seen in our \Lya\ profile is likely associated with the same ISM
inferred from Shapley et al. (2006)\nocite{2006ApJ...651..688S}
(i.e. at the same redshift of the galaxy).

\begin{table*}
\begin{center}
\begin{tabular}{l|c|c|c|c|c|c|c|c|c|c|c|c|c|c|c}
\hline\hline
  &\multicolumn{3}{c}{Single Gaussian} & &\multicolumn{5}{c}{Double Gaussian} & & \multicolumn{5}{c}{Gauss + Absorber} \\
Field &   $z$ & $\sigma$  & $\chi^2$ & & $z_1$ & $\sigma_1$ & $z_2$ & $\sigma_2$  & $\chi^2$ & & $z_G$ &$\sigma_G$ & $z_A$ & EW$_A$ &  $\chi^2$ \\

 & & km s$^{-1}$ & & & & km s$^{-1}$ & & km s$^{-1}$ & & & & km s$^{-1}$  &  & km s$^{-1}$ &  \\

\hline

C15  & 3.098 & 280 & 22 & &
    3.098 & 270 & 3.089 & 109 & 21 & & 
    3.099 & 294 & 3.109 & 379 & 22\\

R1   & 3.100 & 356 & 37 & & 
    3.096 & 137 & 3.103 & 251 & 20 & & 
    3.100 & 385 & 3.090 & 100 & 29 \\

R2  & 3.099 & 409 & 23 & & 
    3.093 & 216 & 3.101 & 302 & 17 & & 
    3.099 & 386 & 3.098 & 20  & 18 \\

R3  & 3.100 & 464 & 56 & & 
    3.092 & 236 & 3.102 & 292 & 21 & & 
    3.099 & 399 & 3.097 & 275 & 22 \\

C11 & 3.099 & 476 & 56 & & 
    3.101 & 108 & 3.098 & 545 & 33 & & 
    3.098 & 422 & 3.095 & 280 & 38\\
 
\hline

\end{tabular}
\caption{Parameters of the fitted profiles (see
Figure~\ref{fig:gausses}) to the \Lya\ emission lines in selected
regions in LAB1. For the Gaussian profiles we show redshift $z$ and
velocity dispersion $\sigma$, and for the Voigt absorber redshift $z$,
and effective width EW. The column density $n$ was fixed to $3 \times
10^{14}$ cm$^{-2}$ for each region. The instrumental dispersion
($\sigma$ = 108 km s$^{-1}$) has been taken into account.}
\label{tab:gausses}
\end{center}
\end{table*}

\subsection{Kinematic signatures of outflow or rotation}
\label{sec:kin}

\begin{figure*}
\begin{center}
\begin{tabular}{c|c|c}

\multicolumn{3}{c}{\psfig{figure=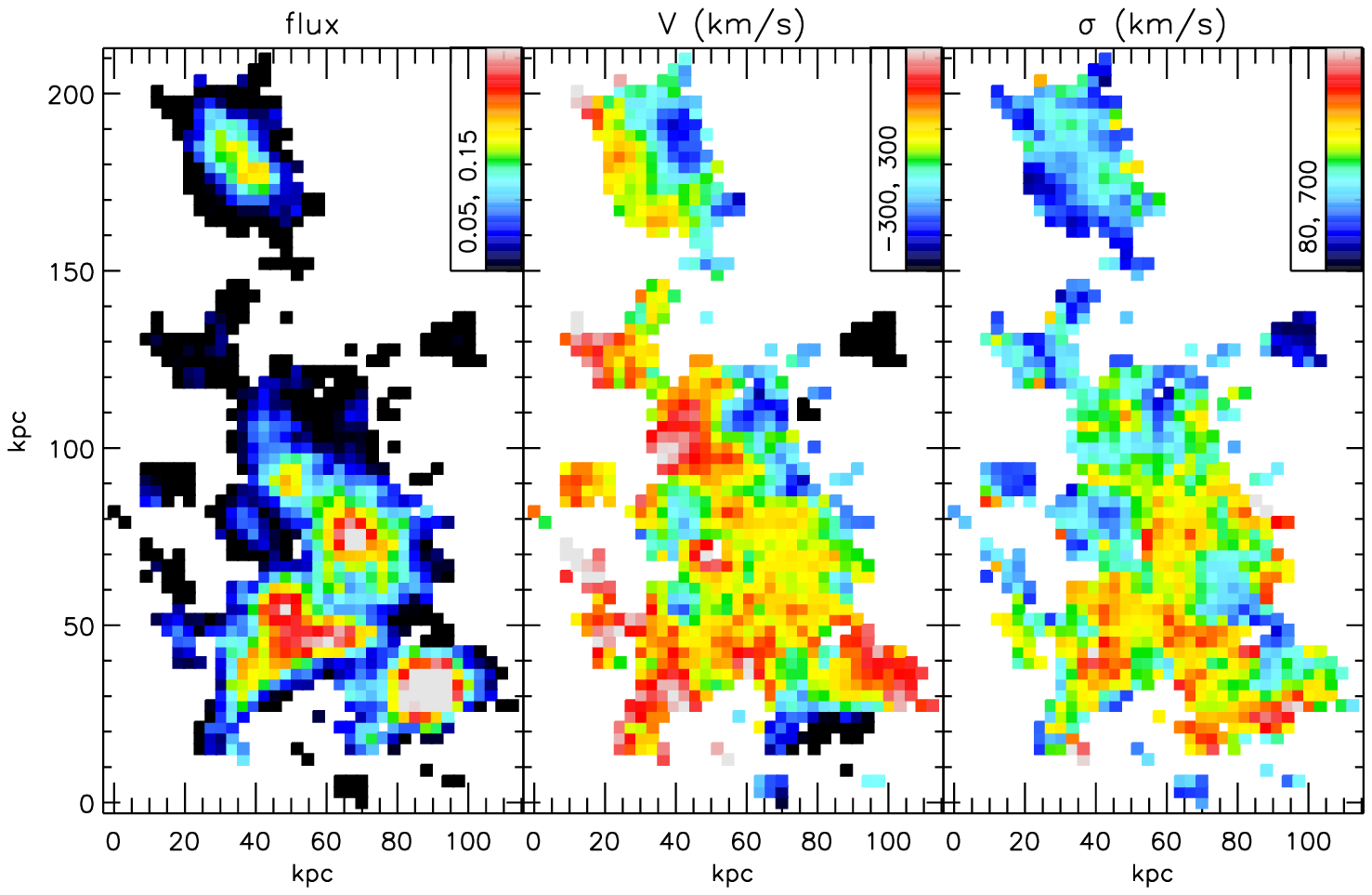,angle=0,width=12.0cm}} \\ 

 & \psfig{figure=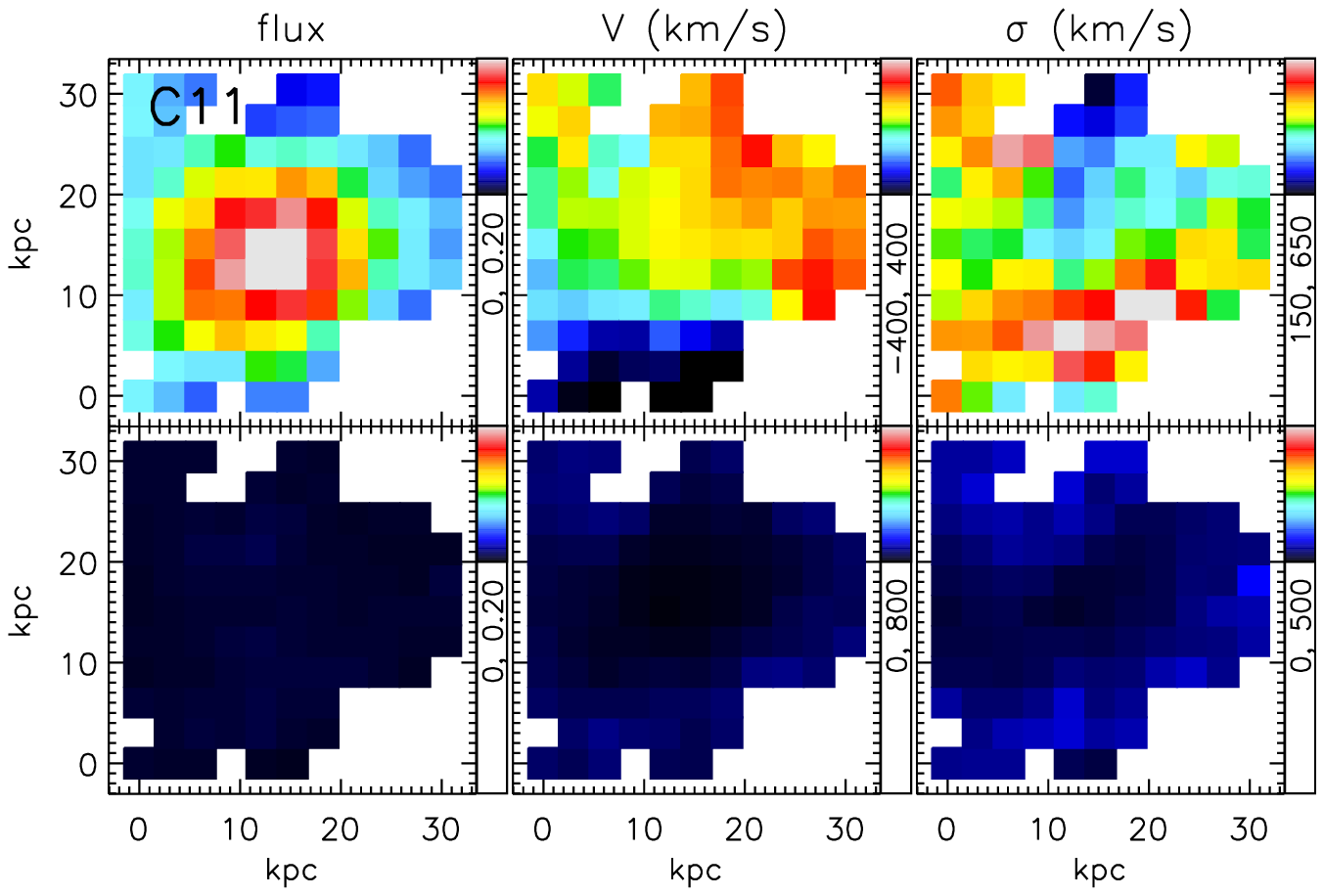,angle=0,width=5.6cm} &
\psfig{figure=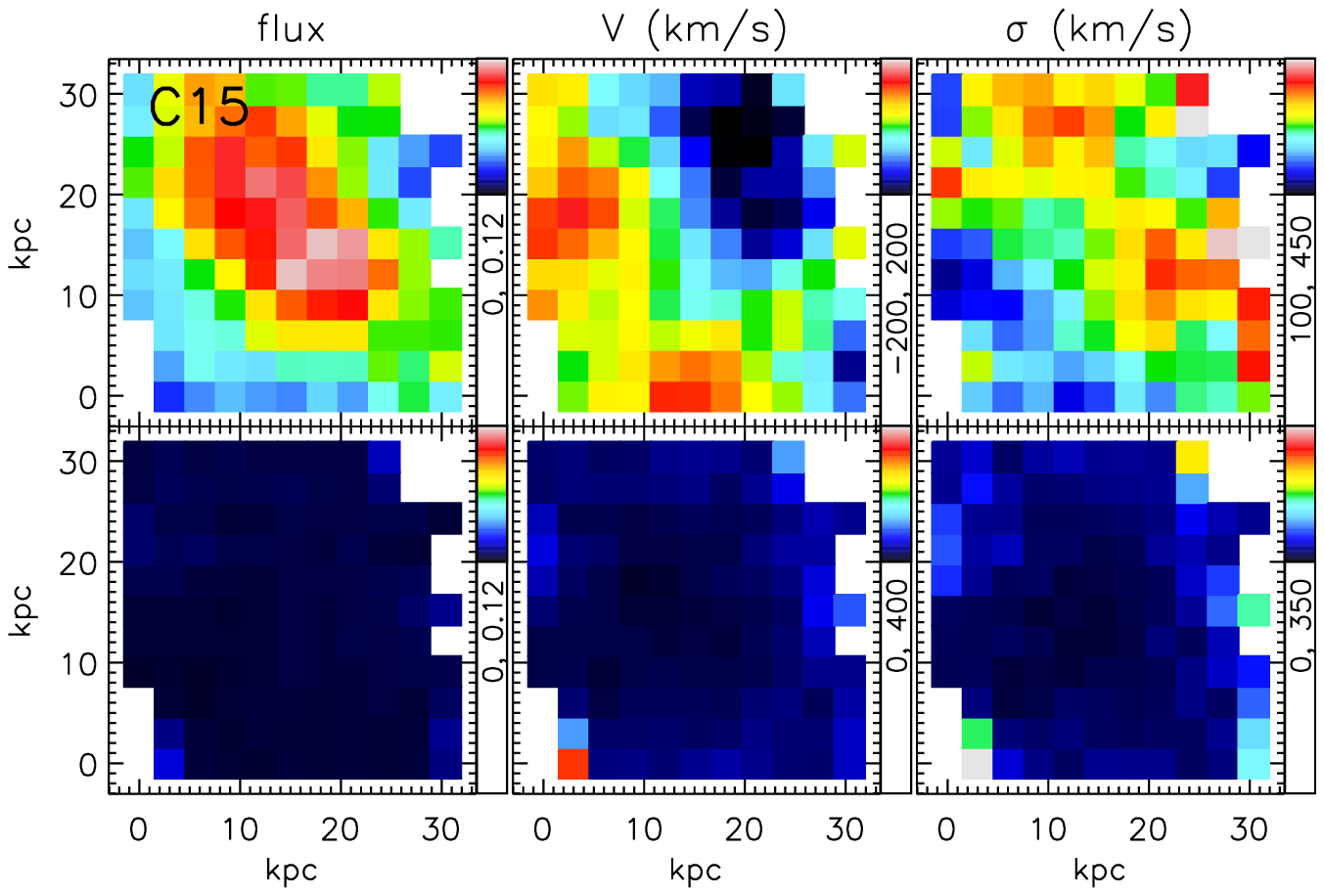,angle=0,width=5.6cm} \\
  
\psfig{figure=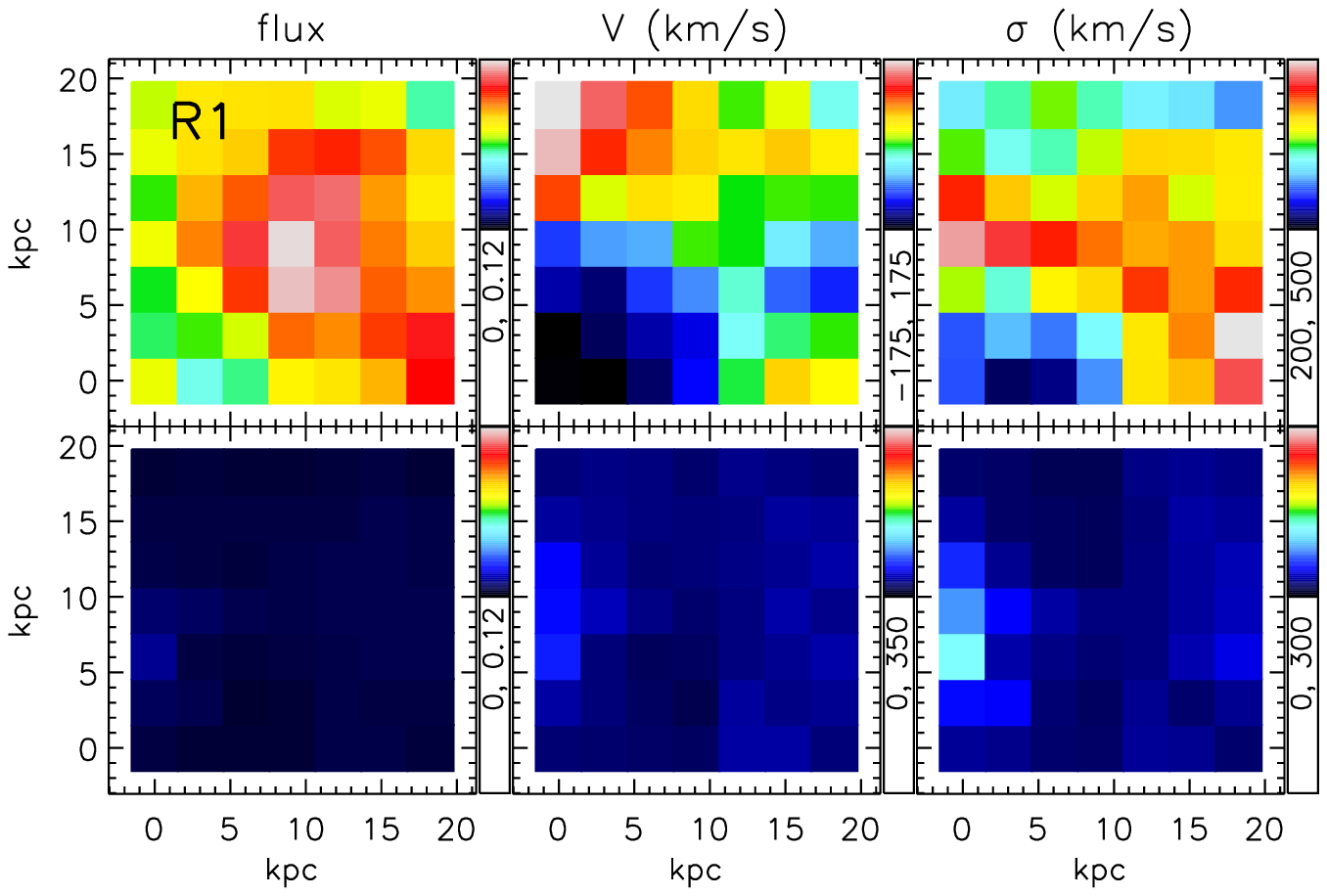,angle=0,width=5.6cm} &
\psfig{figure=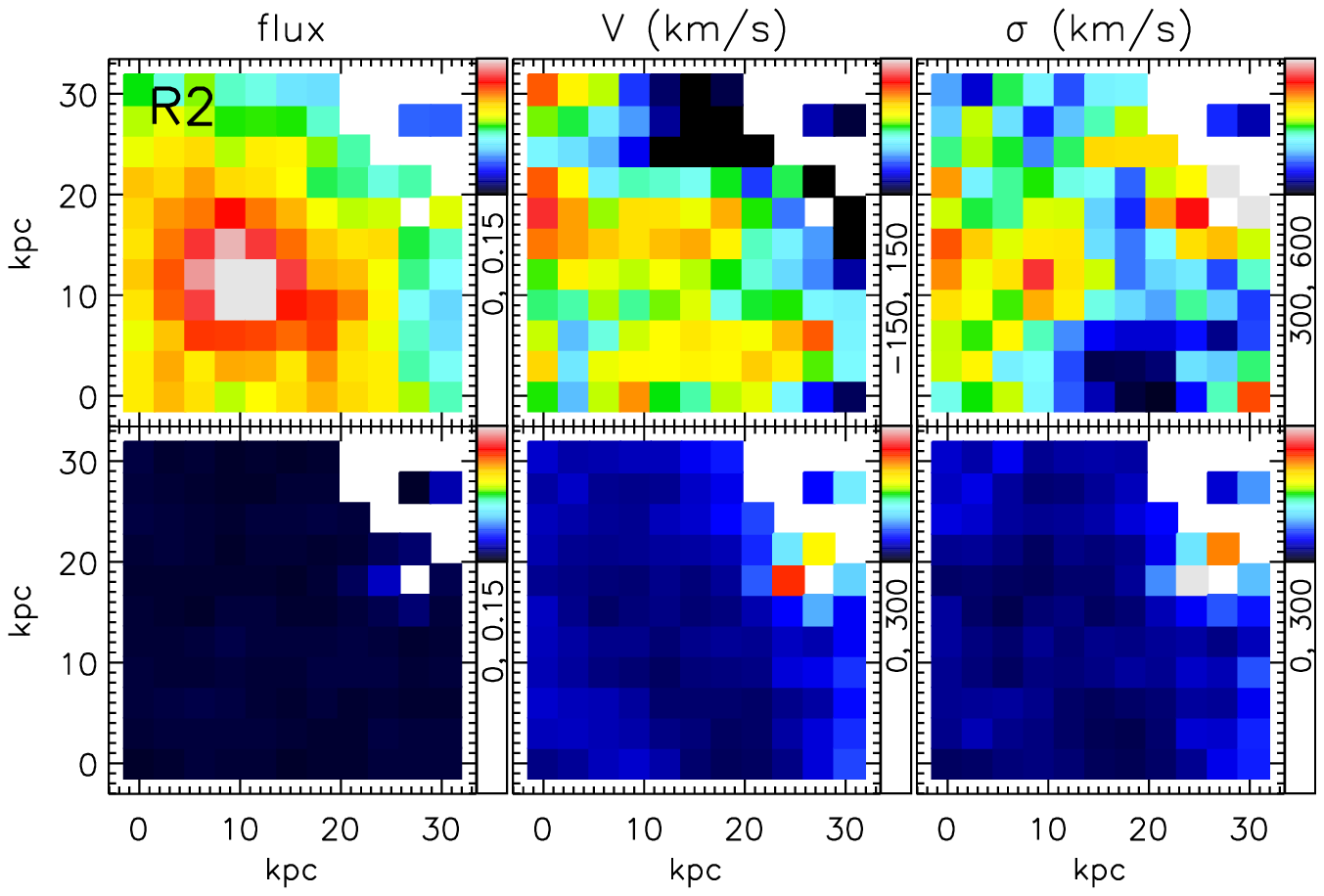,angle=0,width=5.6cm} &
\psfig{figure=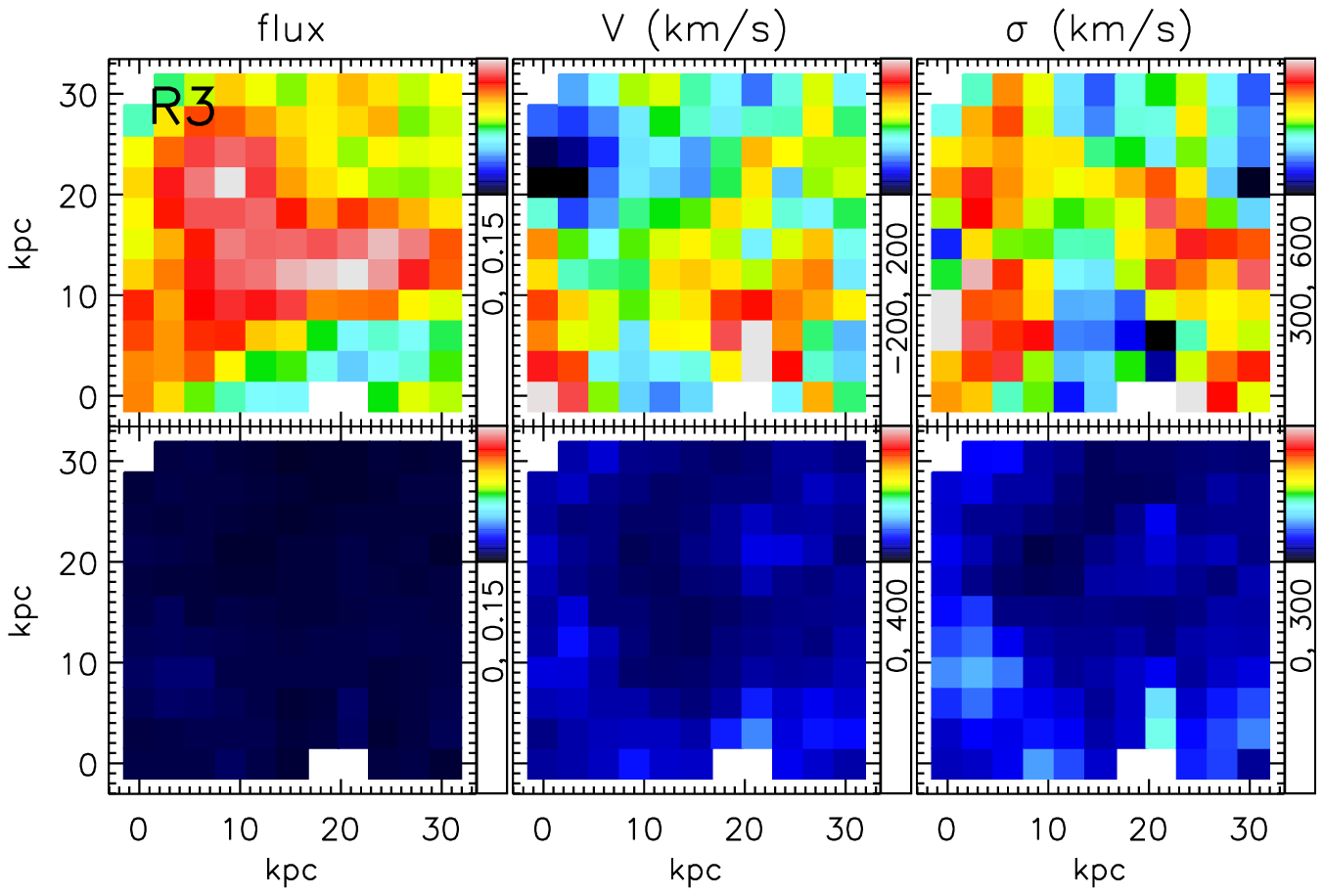,angle=0,width=5.6cm}   \\

\end{tabular}
\caption{Kinematic maps of the \Lya\ gas in LAB1. Top figure shows the
total LAB1 region, with from left to right: flux (in arbritary units),
velocity (km s$^{-1}$) and velocity dispersion (km s$^{-1}$), obtained
by fitting a single Gaussian line to each separate SAURON
spectrum. The colour scale is indicated in the upper right corner of
each plot, and only lines with amplitude-to-noise $A/N > 3$ are
shown. In the remaining figures we show blow-ups of C11 and C15
(middle row, from left to right) and R1, R2 and R3 (bottom row, from
left to right). Top figures of each panel show the kinematic maps
(flux, velocity and velocity dispersion), while the bottom figures
show on the same colour scale the corresponding error (1$\sigma$)
maps, obtained by Monte Carlo simulations.}
\label{fig:vel}
\end{center}
\end{figure*}

Given that the line profiles are generally better matched by a more
complex profile, we investigated the three-dimensional structure of
the emission in each of the emission regions. Since we find no strong
evidence for absorption, it is likely that the line profiles arise
from velocity shear in the emission surrounding each of the
systems. In the discussion that follows we will implicitly assume that
the shift occurs as a response to the bulk velocity of the emitting
gas. The reader should be aware, however, that the line profile has a
complex interaction with the gas velocity field as a result of
radiative transfer effects (Neufeld 1991\nocite{1991ApJ...370L..85N};
Hansen \& Oh 2006\nocite{2006MNRAS.367..979H}; Verhamme, Schaerer \&
Maselli 2006\nocite{2006A&A...460..397V}).

In Figure~\ref{fig:vel} we show kinematic maps of each of the
blobs. At each pixel, we fit a single Gaussian emission line to each
SAURON spectrum. We show only those lines which have an
amplitude-to-noise{\footnote{We define amplitude-to-noise as the ratio
between the amplitude of the fitted Gaussian emission peak and the
noise level of the emission free part of the spectrum.}  $A/N >
3$. The median velocity of the observed field has been subtracted, in
order to reveal any velocity shear that may be indicative of outflow
or rotation. We also show the velocity dispersion, corrected for the
instrumental dispersion. One spatial sampling element is 0.4 $\times$
0.4 arcsec$^2$.

Bower et al. (2004)\nocite{2004MNRAS.351...63B} identified a velocity
shear in the C15 system, and this is confirmed in the deeper data. The
system has a peak to peak shear of $\sim 250$ km s$^{-1}$.  The velocity
field does not have sufficient spatial resolution to distinguish
rotation from outflow (or even inflow). However, the orientation of
the shear, perpendicular to the axis of the underlying galaxy, is
strongly suggestive of outflows such as those seen in the local
starburst galaxy M82 (Shopbell \& Bland-Hawthorn
1998\nocite{1998ApJ...493..129S}; Walter, Weiss \& Scoville
2002\nocite{2002ApJ...580L..21W}).

\begin{figure}
\centerline{\psfig{figure=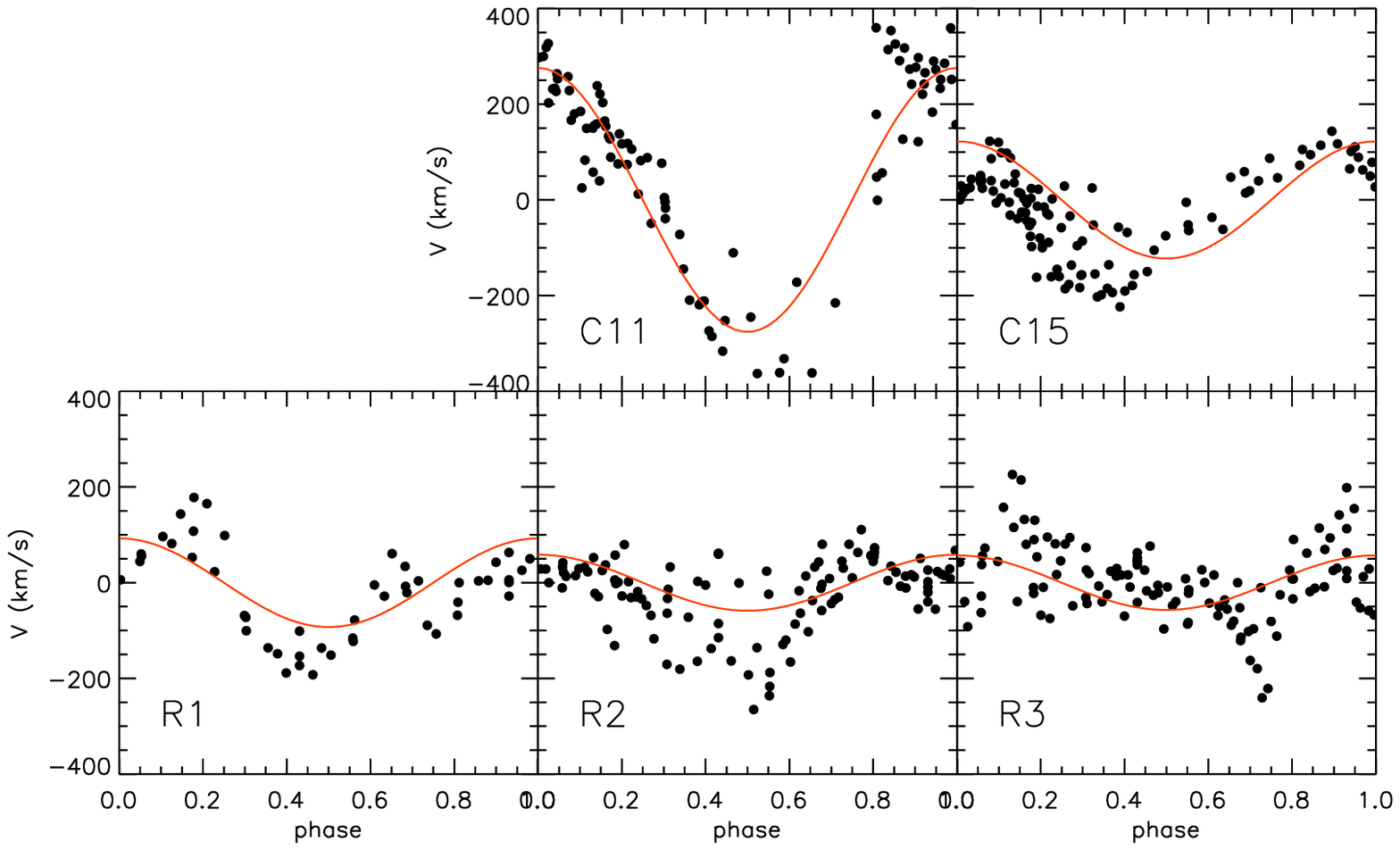,angle=0,width=8cm}}
\caption{Rotation signatures in the \Lya\ regions identified in
LAB1. With black dots we show the observed velocities, and the red
solid line represents the best cosine fit to the data (see text and
Equation~\ref{eq:rot}). The regions C11, C15 and R1-R3 are indicated
in the lower-left corner for each plot. C11, C15 and possibly R1 show
velocity shear signatures, while R2 and R3 do not.}
\label{fig:rotcurves}
\end{figure}

A similar velocity field (but with a substantially higher peak-to-peak
shear of 550 km s$^{-1}$) is revealed for LBG C11, although in this
case the optical source has no clearly identified axis which can help
distinguish between rotation and outflow (or inflow). Weaker evidence
for velocity shear is also appearent in R1 (peak-to-peak velocity
shear of $\sim300$ km s$^{-1}$), while R2 and R3 show no detectable
shear. This is also illustrated in Figure~\ref{fig:rotcurves}, where
we show fits of the observed velocity field ($V_{\mathrm{obs}}$) with
a simple rotation description given by

\begin{equation}
V_{\mathrm{obs}} = V_{\mathrm{rot}} \cos (\phi - \mathrm{PA}).
\label{eq:rot}
\end{equation}

\noindent
Here $V_{\mathrm{rot}}$ is the amplitude of the rotation and PA the
kinematic position angle. The azimuthal angle $\phi$ was defined in
the standard way with respect to a central pixel ($x_c, y_c$). For C11
and C15 we were able to fit for this central pixel, but for R1-R3 the
fit was noisier, and instead we fixed $x_c$ and $y_c$ to the
geometrical centre of the field, which coincides with the peak of
the \Lya\ emission. These fits show more quantitatively than the
velocity maps the absence of a shear pattern in R2 and R3, while C11
and C15 (and to a smaller extent also R1) have clear signatures of
velocity shears. If rotation were indeed responsible for the
observed line widths, then we infer dynamical masses for the separate
blobs of 8$\times 10^{11}$ - 2$\times 10^{12}$ M$_\odot$, in
agreement with the findings of Matsuda et
al. (2006)\nocite{2006ApJ...640L.123M}. If however outflows are the
main driver for the line widths, then the corresponding timescales
vary between 6$\times 10^7$ and 1$\times$$10^8$ yr, also in agreement with
Matsuda et al. (2006)\nocite{2006ApJ...640L.123M}.

In Figure~\ref{fig:spec_ind} we show spectra of individual
  spatial elements for the central parts of the emission regions C15
  and R3. Also indicated for each spectrum is the centre of the single
  Gaussian fit to the total spectrum of these
  regions (see Figure~\ref{fig:gausses}). The velocity shear in the NE-SW direction is clearly
  visible in C15. Note however that most of the flux in C15 is aligned
  perpendicular to this axis, so that the summed spectrum in
  Figure~\ref{fig:gausses} indeed is well approximated by a single
  Gaussian line. In R3 the line profiles and flux distribution is
  however more complex.

Although it is difficult to discern trends on the basis of so few
objects (C11 and C15), it seems that outflows are common in the
systems where an unobscured galaxy is associated with the source of
emission. A much larger sample is needed to draw any quantitative conclusions.

\begin{figure*}
\begin{center}
\begin{tabular}{cc}
\psfig{figure=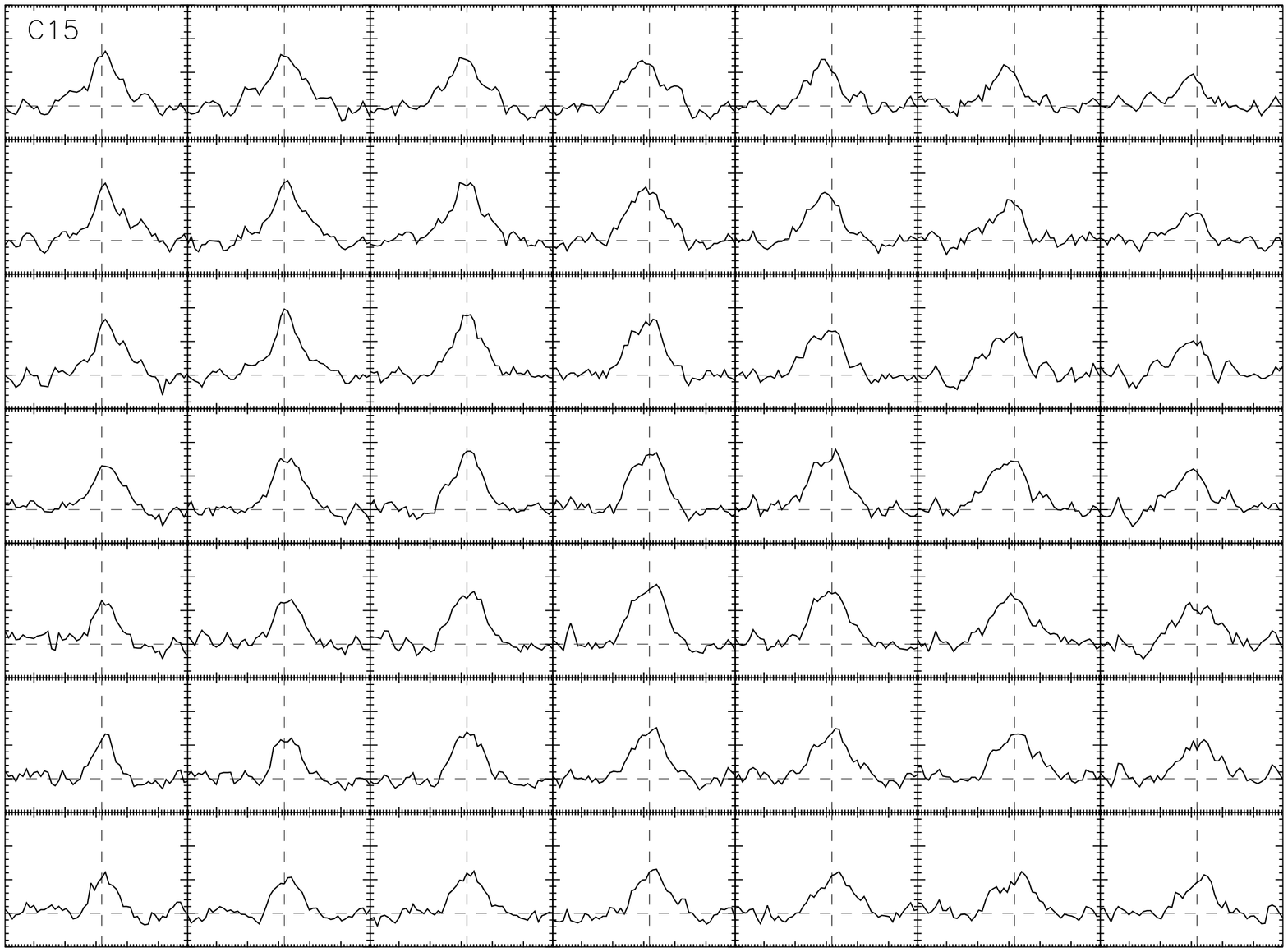,angle=0,width=7.8cm,angle=90} &
\psfig{figure=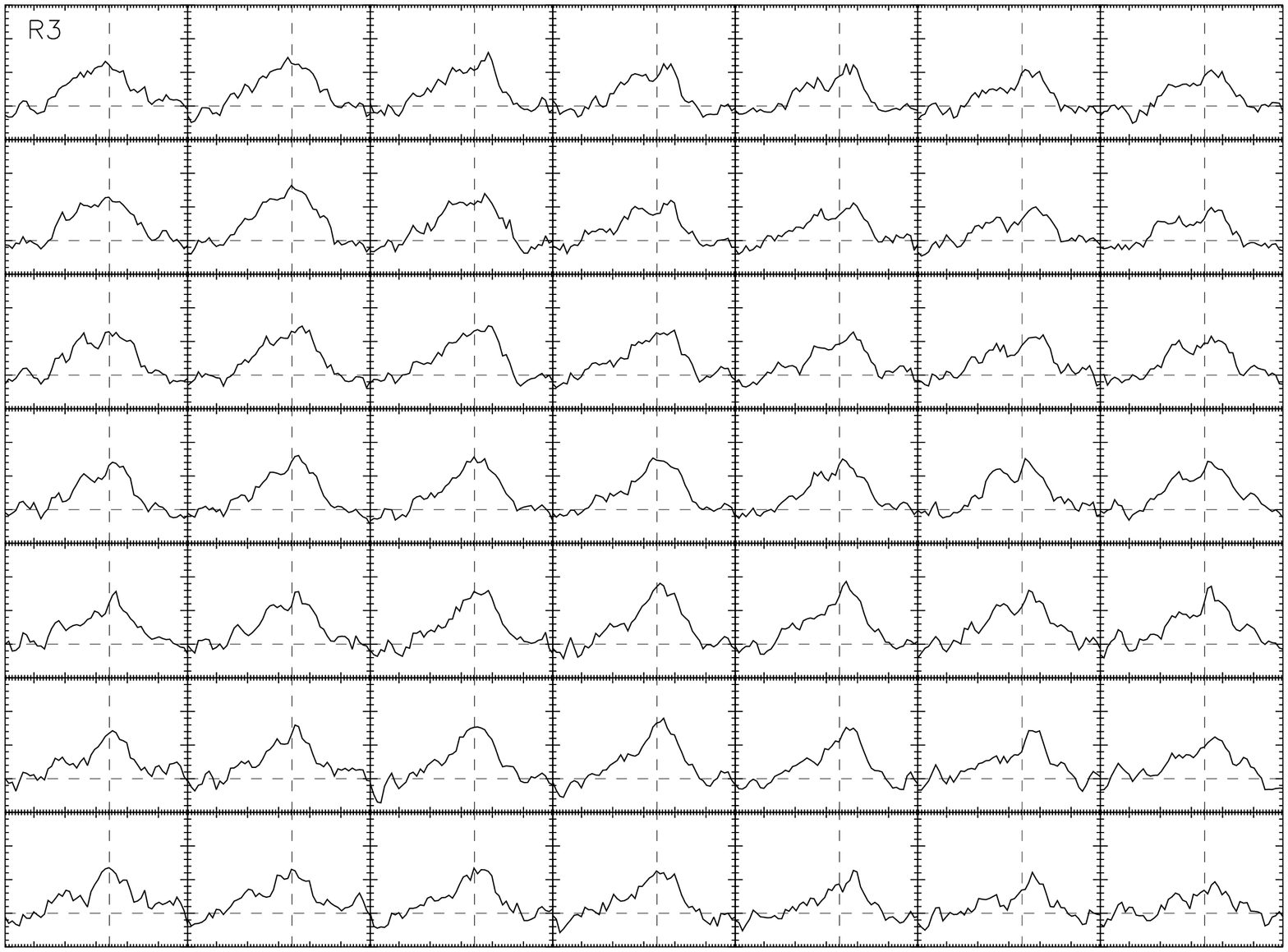,angle=0,width=7.8cm,angle=90}   \\
\end{tabular}
\caption{Spectra of individual spatial elements (0.4 arcsec) of the
central regions of C15 (left) and R3 (right). We show the central 49
spectra over a wavelength range centred on the \Lya\ line (4950 to
5010 \AA). The vertical line is the same in each panel and denotes the centre of the single Gaussian
fit to the summed spectrum in the region (see Figure~\ref{fig:gausses}).}
\label{fig:spec_ind}
\end{center}
\end{figure*}

Interestingly, Shapiro et al. (2008)\nocite{2008ApJ...682..231S}
analysed velocity and dispersion maps of starforming galaxies around
$z \sim 2$, using the \Ha\ emission line. They found that several of
their galaxies could be described by rotating discs. \Lya\ velocity
maps are more difficult to interpret, as \Lya\ is a resonant line and
traces the surface of last scattering. While the \Ha\ emission line
traces the ionised gas in the galaxy disc, \Lya\ lines can also probe
the large-scale diffuse gas structure. Thus it is likely that the two
emission lines probe different gas regimes, and are highly
complementary for studying the formation of young galaxies.

\section{Discussion and Conclusions}
\label{sec:conclusion}

We have presented very deep integral-field spectrograph observations of 
the LAB1 \Lya\ emission halo. The deep data allow us to study the spatial and 
velocity structure of this system in unprecedented detail.

We find that the giant halo is made up from a superposition of five
distinct blobs, with the emission from these regions accounting for at
least 55 per cent of the total diffuse flux. Most of the emission
regions are associated with individual galaxies. The regions C15 and
C11 are associated with optical Lyman Break Galaxies and region R3 is
associated with a bright submillimeter source. While this source is
faint in the optical, it is bright at 3.6 $\mu$m, suggesting that it
is a strongly dust-obscured starforming galaxy. None of these sources
are detected at X-ray wavelengths, suggesting that it is unlikely that
the emission is powered by an AGN. The regions R1 and R2 are not
associated with any optical or IR source, and the emission from these
blobs plausibly comes from gas associated with the proto-cluster
potential.

The integral-field spectra allow us to examine the emission line
profiles and velocity structure in each of the blobs. In R1, R3 and
C11, we find that the integrated emission line profiles are not
adequately fitted by a single component Gaussian, and that a better
fit is obtained with either a two component Gaussian or the
combination of a Gaussian emission and Voigt absorption line
profile. We find that the
underlying emission has a velocity centroid that is extremely similar
from region to region, reinforcing the idea that the LAB1 system is a
virialised group with velocity dispersion $\sim100$ km s$^{-1}$.
However, in contrast to the LAB2 system, we find no evidence for a
coherent shell of absorption that covers the entire system. Any
absorption features are significantly weaker than those seen in LAB2,
so that they might arise in the large-scale structure foreground to
the proto-cluster.

C15, C11 and possibly R1 show evidence of coherent velocity shear
arising from an outflow or rotation. In C15 the velocity gradient is
perpendicular to the morphology of the underlying galaxy, consistent
with the pattern expected for an outflowing galactic wind. In the
other systems the relation between the velocity field and the
underlying galaxy is unclear. The velocity shear is largest in C11
where it is $\sim 550$ km s$^{-1}$ over approximately 25 kpc.  The
implied outflow velocity is comparable with that seen in many other
Lyman break systems, and does not suggest that the sources seen in
LAB1 are undergoing unusually strong feedback.

The primary motivation for these observations was to discover whether
the coherent absorption systems such as seen by Wilman et
al. (2005)\nocite{2005Natur.436..227W} in LAB2 are a ubiquitous
feature. These observations have shown that they are not.  The data
for LAB2 are best interpreted as a large-scale super-bubble of
material that has been expelled by a high power, perhaps explosive
feedback event. The two LAB systems both seem to be made up of smaller
emission clouds. So why does the absorption pattern in LAB1 and LAB2
differ? One possible answer is that a similar event has occured in
LAB1 in the past, but that the shell has now broken up, as it
  passed through the intergalactic medium of this proto-cluster, or that
it is sufficiently blue shifted that it cannot be seen in absorption
against the system's \Lya\ emission. Another explanation might be that
LAB1 is a younger system in which the large scale outflow is yet to
develop. This seems unlikely, however, since we see no signs of
spectacular outflows associated with the individual emission systems.
Now that we have dissected the giant emission halo into a number of
smaller systems, these seem quite comparable to the many radio quiet
LAB systems identified by Matsuda et
al. (2004\nocite{2004AJ....128..569M}). In many ways, the composite
system represents a microcosm of diffuse \Lya\ emission in general,
with the systems reflecting a diversity of power sources. Geach et al.
(2009)\nocite{2009ApJ...700....1G} favour photo-ionisation as the
principle power source for LABs. This fits in well with C11, C15 and
R3 (if we allow for the possibility that it is only so strongly
obscured along our line of sight). However, this appears to describe
the R1/R2 emission less well. Possibly the emission from this part of
the system is much more closely related to the emission seen around
radio galaxies (Chambers et al. 1990\nocite{1990ApJ...363...21C};
Villar-Mart\'in et al. 2002\nocite{2002MNRAS.336..436V}) and (perhaps)
in local cooling flow clusters (Johnstone \& Fabian
1988\nocite{1988MNRAS.233..581J}).


\section*{Acknowledgements}

We would like to thank the anonymous referee for comments and
suggestions, which helped to improve the presentation of our work. It
is a pleasure to thank Michele Cappellari, Eveline van Scherpenzeel,
Chris Benn and the ING staff for support on La Palma. We gratefully
acknowledge Huub R\"ottgering, Joop Schaye, Kristen Shapiro and Ian
Smail for fruitful discussions, and Roland Bacon and Eric Emsellem for
their help in the initial stages of this project.

This research was supported by the Netherlands Research School for
Astronomy NOVA, and by the Netherlands Organization of Scientific
Research (NWO) through grant 614.000.426. AW acknowledges The Leids
Kerkhoven-Bosscha Fonds and the European Southern Observatory for
contributing to working visits. AMS and JEG acknowledge support from
STFC.

The SAURON observations were obtained at the William Herschel
Telescope, operated by the Isaac Newton Group in the Spanish
Observatorio del Roque de los Muchachos of the Instituto de
Astrof\'isica de Canarias.



\label{lastpage}
\end{document}